\documentclass[a4paper,11pt]{article}
\usepackage{jheppub}
\usepackage{amssymb}

\usepackage[utf8]{inputenc}
\usepackage[english]{babel}
\usepackage{mathtools}
\usepackage{amsfonts}
\usepackage{amsthm} 
\usepackage{physics} 
\usepackage{comment} 
\usepackage[dvipsnames]{xcolor}
\usepackage{float} 
\usepackage{tikz} 
\usetikzlibrary{shapes,arrows,positioning,automata,backgrounds,calc,er,patterns}
\usepackage[compat=1.0.0]{tikz-feynman}
\usepackage{graphicx}
\usepackage{tabularx}
\usepackage{amssymb}
\usepackage{mathrsfs}

\usepackage{bbm}

\renewcommand{\d}{\delta}

\renewcommand{\th}{\theta}

\newcommand{\w}{\omega}
\newcommand{\G}{\Gamma}

\newcommand{\N}{\mathbb{N}} 
\newcommand{\Z}{\mathbb{Z}} 

\newcommand{\DeclareAutoPairedDelimiter}[3]{%
  \expandafter\DeclarePairedDelimiter\csname Auto\string#1\endcsname{#2}{#3}%
  \begingroup\edef\x{\endgroup
    \noexpand\DeclareRobustCommand{\noexpand#1}{%
      \expandafter\noexpand\csname Auto\string#1\endcsname*}}%
  \x}
\DeclareAutoPairedDelimiter{\p}{(}{)} 

\renewcommand{\bar}[1]{\mkern1mu\overline{\mkern-1mu#1\mkern-1mu}\mkern1mu}
\renewcommand{\tilde}[1]{\widetilde{#1}}

\title{Spinning Particle Geometries in AdS$_3$/CFT$_2$}

    \author{Ziyi Li }
\affiliation{Department of Physics, University of California at Davis, California 95616, USA
}

\emailAdd{zzyli@ucdavis.edu}

\abstract{We study spinning particle/defect geometries in the context of AdS$_3$/CFT$_2$. These solutions lie below the BTZ threshold, and can be obtained from identifications of AdS$_3$. We construct the Feynman propagator by solving the bulk equation of motion in the spinning particle geometry, summing over the modes of the fields and passing to the boundary. The quantization of the scalar fields becomes challenging when confined to the regions that are causally well-behaved. If the region containing closed timelike curves (CTCs) is included, the normalization of the scalar fields enjoys an analytical simplification and the propagator can be expressed as an infinite sum over image geodesics. In the dual CFT$_2$, the propagator can be recast as the HHLL four-point function, where by taking into account the $PSL \p{2,\mathbb{Z}}$ modular images, we recover the bulk computation. We comment on the casual behavior of bulk geometries associated with single-trace operators of spin scaling with the central charge below the BTZ threshold.
}

\begin{document} 
\maketitle
\flushbottom

\section{Introduction}
The AdS/CFT correspondence \cite{Maldacena:1997re} establishes an equivalence between quantum gravity in Anti-de Sitter space and strongly coupled gauge theories in one less dimension. Observables of the gauge theory such as the correlation functions can be computed in the gravity side as boundary correlators \cite{Witten:1998qj,Gubser:1998bc}. In the strongly coupled limit, the quantum gravity theory is often realized as perturbative string theories, where semi-classical geometries can be trusted. In particular, black holes in the bulk are dual to thermal states in the field theory \cite{Maldacena:2001kr,Grinberg:2020fdj,Berenstein:2022nlj}. Much progress has been made to make these statements more precise, especially in lower dimensions where analytical properties on both sides can be extrapolated. AdS$_3$/CFT$_2$ provides a robust avenue to test the duality while avoiding some of the computational difficulties. Since gravity in three dimensions has no local propagating degrees of freedom \cite{Giddings:1983es,Deser:1983tn,Deser:1983nh}, many geometries in AdS$_3$ can be obtained from identifications, including uncharged black holes. These solutions, known as BTZ geometries, share many features of higher dimensional counterparts, yet many computations can be made exact \cite{Banados:1992wn,Banados:1992gq,Carlip:1995qv}. Below the BTZ threshold, these geometries contain conical singularities, which can be interpreted as heavy point particles that cause the geometry to backreact. These solutions appear to play an important role in the AdS$_3$/CFT$_2$ correspondence. Classical collision processes of these particles can produce black holes if the total energy is above certain threshold \cite{Gott:1990zr,Steif:1995zm,Matschull:1998rv,Holst:1999tc,Balasubramanian:1999zv,Lindgren:2015fum,Haehl:2023lfo}. Moreover, by studying the limit where the particle number goes to infinity, one recovers the usual thin-shell collapse analogous to higher dimensions \cite{Lindgren:2016wtw,Lindgren:2017hiu}, and the CFT$_2$ dual was studied in \cite{Anous:2016kss} using the monodromy method. The defect solutions have also appeared in the replica trick for gravitational entropy \cite{Lewkowycz:2013nqa}. More interestingly, they have also appeared in the studies of 3D gravity partition functions \cite{Maloney:2007ud,Keller:2014xba}, where by adding these heavy particle states, it was shown that unitarity could be restored explicitly for the partition function \cite{Alday:2019vdr,Benjamin:2020mfz}. More recently, a unitary partition function in 3D gravity was constructed, where by adding two states with spin scaling with the central charge $c$, unitarity is restored in the large spin limit \cite{DiUbaldo:2023hkc}. These states were interpreted as strongly coupled spinning strings in AdS$_3$ found by Maxfield and Wang \cite{Maxfield:2022rry}. The spacetime metric outside the spinning strings takes the same form as the spinning particle of the BTZ solutions \cite{Miskovic:2009uz,Casals:2016ioo,Casals:2016odj,Martinez:2019nor,Briceno:2021dpi}. In this note, we will mainly focus on the non-extremal spinning particles, and study the boundary propagator in this background. The method we employ for computing the propagator was first introduced in the context of real-time holography \cite{Skenderis:2008dh,Skenderis:2008dg}, which uses standard techniques of quantum field theory in curved spacetime, and it has later been extended to the investigation below the BTZ threshold \cite{Arefeva:2016wek,Hijano:2019qmi,Berenstein:2022ico}. 

\subsection{Outline}
We begin in \S.\ref{section2} with a review of the spectrum of the family of BTZ solutions, and demonstrate how the spinning particle geometry can be obtained by identification from empty AdS$_3$. In \S.\ref{section3}, focusing on the non-extremal spinning particle solution, we solve the scalar's equation of motion, extrapolate the quantization condition, and compute the normalization constant. We show that when the region containing closed timelike curves (CTCs) is included, there is a natural boundary condition to impose in the bulk, and the normalization constant enjoys an algebraic simplification. The scalar fields can then be canonically quantized and used to construct the Feynman propagator. In \S.\ref{Section4}, we construct the boundary propagator in the spinning geometry containing CTCs, where we show that the full propagator can be expressed as an infinite sum over image geodesics up to appropriate regularization. The same infinite sum over image geodesics can similarly be applied to the non-spinning defect with non-integer values of angle deficit. We study our results from the dual CFT in \S.\ref{section5}, where we show that the HHLL four-point function of Fitzpatrick, Kaplan, and Walters \cite{Fitzpatrick:2015zha} produces the leading divergence of the boundary propagator computed in the bulk. The subleading contributions can also be matched if we follow the proposal \cite{Maloney:2016kee} to take into account the $PSL \p{2,\mathbb{Z}}$ modular images, which are precisely the sum over image geodesics in the bulk. This can be viewed as taking into account the vacuum blocks across
all channels and summing over all of them. We also discuss the implications of the casual behavior of the bulk geometry that is dual to this state. We end with a discussion and some future directions in \S.\ref{section6}.
\section{Family of BTZ solutions}
\label{section2}
\begin{figure}[h]
    \centering
    \hspace*{-2cm}
    \includegraphics[width=1.4\textwidth]{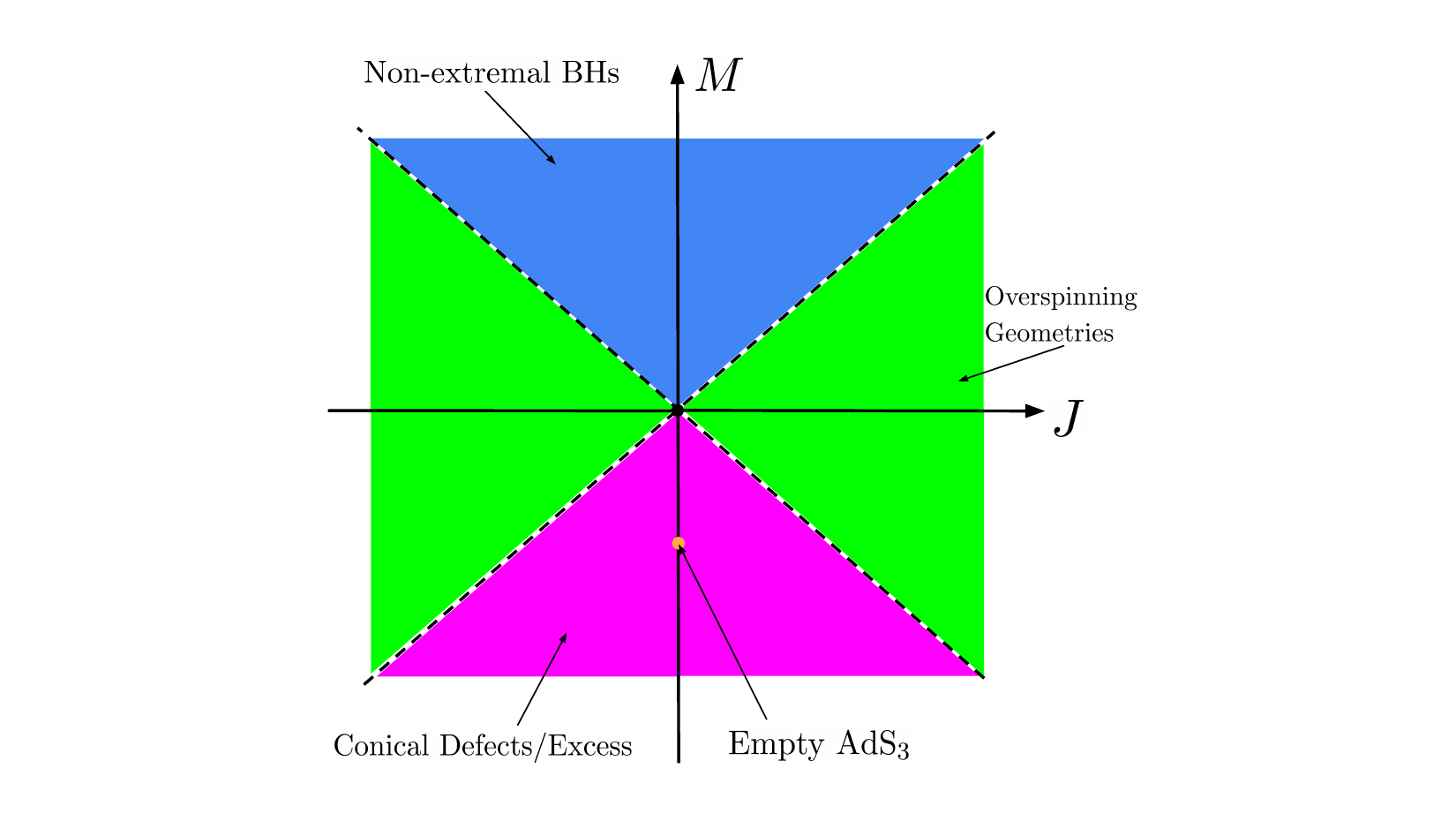}
    \caption{$M-J$ graph that represents the solutions of the BTZ metric. The blue region represents the non-extremal black hole solution with $M>\abs{J}>0$. The green region represents the overspinning geometries with $\abs{M}<\abs{J}$. The magenta region represents conical defect/excess, which is separated by $M^2-J^2=1$. The black dot at the origin is the massless BTZ geometry with $M=J=0$, and the orange dot below is the empty AdS$_3$ geometry with $M=-1$. The dashed lines represent extremal BH/Defect solutions. We will mainly focus on the magenta region with angular deficit.   }
    \label{fig:geometries}
\end{figure}

In (2+1)-dimensional Einstein gravity, the curvature tensor is completely determined by the Ricci tensor and solutions of the vacuum Einstein's equation with a cosmological constant $\Lambda$,
\begin{align}
    R_{\mu\nu}=2\Lambda g_{\mu\nu}
\end{align}
have a constant curvature \cite{Deser:1983nh}. This includes the family of BTZ solutions \cite{Banados:1992wn,Banados:1992gq,Carlip:1995qv} for negative values of the cosmological constant, which is given by the line element: 
\begin{equation}
\label{metric}
    \dd s^2=-\p{\frac{r^2}{l^2}-M}\dd t^2+\p{-M+\frac{r^2}{l^2}+\frac{J^2}{4r^2}}^{-1}\dd r^2+r^2 \dd \phi^2-J\dd \phi \hspace{1mm} \dd t
\end{equation}
where $l^2=-1/\Lambda$\footnote{We will set $l=8G=1$ from now on, and restore them when necessary.}, and $M,J$ are the mass and angular momentum of the spacetime respectively. The BTZ solutions can be obtained as a quotient of the universal covering space $\tilde{\text{AdS}}$ by different group isometries. From a geometric perspective, AdS$_3$ spacetime can be constructed from flat $\mathbb{R}^{2,2}$ space, where the isometry algebra of AdS$_3$ is then $SO(2,2) \approx SL(2,\mathbb{R})\times SL(2,\mathbb{R})/\mathbb{Z}_2 $, with group elements $(\rho_L,\rho_R)\sim(-\rho_L,-\rho_R)$. The above metric can be viewed as quotient by the group generated by the isometry: 
\begin{align}
  \rho_L  =\left(
\begin{array}{cc}
 e^{\pi \p{r_+-r_-}} & 0 \\
 0 & e^{-\pi \p{r_+-r_-}} \\
\end{array}
\right), \quad
\rho_R=\left(
\begin{array}{cc}
 e^{\pi \p{r_++r_-}} & b \\
 0 &  e^{-\pi \p{r_++r_-}}\\
\end{array}
\right)
\end{align}
where $r_{\pm}$ are given in (\ref{roots}). Different ranges of $M,J$ correspond to different conjugacy classes of $\rho_{L,R}$, which are nicely summarized in \cite{Martinez:2019nor} (See also Figure.\ref{fig:geometries}):
\begin{itemize}
    \item $M\geq 0$ and \hspace{0.5mm} $J=M$: \hspace{2.4mm}Extremal BTZ geometry (Parabolic).
    \item $M>0$ and $\abs{J} < M$: \hspace{1.8mm}Black holes (Hyperbolic).
    \item $M<0$ and $\abs{J}<\abs{M}$: Spinning particles (Elliptic).
    \item $M<0$ and $\abs{J}=\abs{M}$: Extremal spinning Particles. (Parabolic)
    \item $M<0$ and $\abs{J}>\abs{M}$: Overspinning Particles\footnote{It is not clear to us if these solutions exist in the dual field theory.}
    \item $M=-1$ and $J=0$: \hspace{2.8mm} Empty AdS$_3$
\end{itemize}
Note that for all ranges of these parameters, the spacetime is free of CTCs in the region $r^2> 0$, as is evident from the metric. More fundamentally, this can be understood from the perspective of the covering space. As discussed above, AdS$_3$ space can be obtained from flat $\mathbb{R}^{2,2}$ space: 
\begin{align}
    \dd s^2=-\dd T_1^2-\dd T_2^2+\dd X_1^2+\dd X_2^2
\end{align}
by restricting to the pseudosphere $X^2_1+X^2_2-T_1^2-T_2^2=-1$. The manifold has six Killing vector fields (KVFs) given by the linear combinations of the $SO(2,2)$ generators: 
\begin{align}
    \Theta=\frac{1}{2}\omega^{AB}J_{AB}
\end{align}
where $\omega_{AB}$ is an anti-symmetric tensor and $J_{AB}=2X_{[A}\partial_{B]}$. The six geometries listed above can then be obtained by identifications along the orbits of the six KVFs, and the explicit forms of the KVFs can be found in \cite{Banados:1992gq,Briceno:2021dpi}. The identification procedure joins two points of the covering space that are in the same orbit of the KVF in the quotient space, and one should require the KVF to be spacelike to avoid identifying two points that are time-like separated in the covering space, which will give rise to CTCs in the quotient space. This requirement corresponds to: 
\begin{align}
    \Theta \cdot \Theta >0 \implies r^2>0
\end{align}
which is sufficient to ensure a causally well-behaved quotient geometry. 

In the positive spectrum of the parameter $M$, the geometry is identified with a rotating black hole, and the two event horizons are given by the roots of $g^{rr}$: 
\begin{align}
\label{roots}
    r_{\pm}=\frac{1}{2}\p{\sqrt{M+J}\pm \sqrt{M-J}}
\end{align}
However, in the cases where $M<0$, the roots become imaginary and the geometries become conical singularities. These solutions were first studied in \cite{Miskovic:2009uz} in the context of the supersymmetric extension of 2+1 dimensional gravity. The quantum aspects of these solutions were later investigated in \cite{Casals:2016ioo,Casals:2016odj}, where by studying the backreaction of a quantum scalar field on the geometry, it was found that the back-reacted geometry develops a horizon, thus shielding the conical singularity. More recently, a complete study of the geodesic structures of these spinning particles was done in \cite{Martinez:2019nor,Briceno:2021dpi}. In this note, we will mainly focus on the case of the non-extremal spinning particle.
\subsection{The metric}
The non-extremal spinning particle has $M<0,\abs{J}<\abs{M}$, and we define $-1/n^2=M$, $n \in \mathbb{R}\setminus\{0\}$. The two roots (\ref{roots}) of $g^{rr}$ becomes: 
\begin{align}
      r_{\pm}=i \beta_{\pm}, \quad \beta_{\pm}=\frac{1}{2}\p{\sqrt{\frac{1}{n^2}-J}\pm\sqrt{\frac{1}{n^2}+J}}
\end{align}
where $\beta_{\pm}$ are real. The metric expressed in terms of $\beta_{\pm}$ is then: 
\begin{align}
\label{metric2}
    \dd s^2=-\p{r^2+\beta_+^2+\beta_-^2}\dd t^2+ \frac{r^2 \dd r^2}{\p{r^2+\beta_+^2}\p{r^2+\beta_-^2}}+r^2 \dd \phi^2 +2\beta_{-}\beta_+\dd \phi \hspace{1mm} \dd t
\end{align}
As discussed above, the geometry can be constructed from the identification of empty AdS$_3$:
\begin{align}
\dd s^2=-\p{1+\hat{r}^2}\dd \hat{t}^2+\p{1+\hat{r}^2}^{-1} \dd \hat{r}^2+\hat{r}^2
\dd \hat{\phi}^2 \notag 
\end{align}
Explicitly the identification can be implemented from the following coordinate transformations,
\begin{align}
\label{coortrans}
        \hat{t}&=\beta_-\phi-\beta_+t \notag\\
    \hat{\phi}&=\beta_+\phi -\beta_- t \notag \\
    \hat{r}^2&=\frac{r^2+\beta_{-}^2}{\beta_+^2-\beta_-^2}
\end{align}
and we recover the metric (\ref{metric2}) with the periodicity condition $(\hat{t},\hat{\phi})\sim (\hat{t}+2\pi\beta_-,\hat{\phi}+2\pi \beta_+)$. It is important to note that if we start with empty AdS$_3$, the range of the radial coordinate is:
\begin{align}
    0\leq \hat{r}^2 < \infty
\end{align}
the corresponding range of radial coordinate in the case of a spinning particle is then:
\begin{align}
    -\beta_-^2\leq r^2 < \infty
\end{align}
This means that if we simply perform the above identification in pure AdS$_3$, the resulting spinning geometry will contain the region with CTCs, i.e., the region where $\Theta \cdot \Theta<0$. In the case of the classical geometry, we can simply exclude this region by restricting the coordinate range to be $0\leq r^2 < \infty$, and the classical geometry is casually well-behaved. However, as we shall see, restricting to the causally well-behaved region makes an analytical calculation more cumbersome when studying the quantization of the scalar fields.

The above solutions also include the folded spinning string solution of Maxfield and Wang \cite{Maxfield:2022rry}, where the metric outside the string takes the form of (\ref{metric}) in the $(t,r,\phi)$ coordinates with:
\begin{align}
    M=\frac{\epsilon_L+\epsilon_R}{2}, \quad J=\frac{\epsilon_R-\epsilon_L}{2}
\end{align}
and the string is parameterized with the tension $\lambda$ and angular velocity $\omega$, where the spacetime energy and spin can be solved in terms of. The spinning strings belong to the elliptic conjugacy class since $\epsilon_{L,R}<0$. The string is also free of CTCs if one restricts to the region $r^2>0$\footnote{In \cite{Maxfield:2022rry}, the authors studied the solution in the Fefferman-Graham coordinates, where the radial coordinate $z$ is related to $r$ via $r^2=\frac{1}{z}\p{z+\frac{\epsilon_L}{4}}\p{z+\frac{\epsilon_R}{4}}$. If one assumes that $\abs{\epsilon_R}<\abs{\epsilon_L}$, the region containing CTCs is $z<-\frac{\epsilon_L}{4}$, which corresponds to $r^2<0$.}. The string tension is given by:
\begin{align}
    \lambda= \frac{1}{2\pi}\frac{1}{\ell_s\ell_p}
\end{align}
where $\ell_s,\ell_p$ are the string and Planck scales measured in the AdS scale respectively. For general values of $\lambda, \omega$, the solutions are expressed in terms of elliptic integrals. But at a specific value of $\lambda=1$, the equations simplify and solutions can be obtained in terms of elementary functions. Interestingly, spinning strings have a maximum value of angular momentum, which is characterized by $\epsilon_L=0, \hspace{1mm} \epsilon_R>0$. Note that in this case the mass parameter becomes positive and the string transitions to an extremal rotating BTZ black hole with horizon area of order $\lambda^{-1}$ \cite{Maxfield:2022rry}.

\section{Quantizing the scalar field}
\label{section3}
We now turn to the study of scalar fields in the spinning particle geometry. 
\subsection{Solutions of the wave equation}
Consider a free massive scalar field propagating in this background. The solution is separable due to the existence of two KVFs: $\p{\partial/\partial t}^a,\p{\partial/\partial \phi}^a$. The radial solution was solved in the rotating BTZ black hole \cite{Keski-Vakkuri:1998gmz,Birmingham:2001hc}, and it generalizes trivially to our case. The wave equation reads: 
\begin{align}
   \p{\nabla^2-m^2}\Phi(\mathbf{x})=0, \quad \Phi\p{\mathbf{x}}=R(r)e^{-i\omega t} e^{i \ell \phi}
\end{align}
with the following coordinate transformation:
\begin{align}
\label{cotrans}
    z=\frac{r^2+\beta_{-}^2}{r^2+\beta_{+}^2}
\end{align}
the radial equation is:
\begin{align}
    z(1-z)\frac{d^2R(z)}{dz^2}+(1-z)\frac{dR(z)}{dz}+\p{-\frac{m^2/4}{(1-z)}-\frac{\alpha^2}{z}+\gamma^2}R(z)=0
\end{align}
where
\begin{align}
    \alpha^2&=\frac{\left(\omega\beta _-  -\ell\beta _+ \right)^2}{4 \left(\beta _+^2-\beta _-^2\right)^2}, \\[2pt]
    \gamma^2&=\frac{\left(\ell\beta _- -\omega\beta _+  \right)^2}{4 \left(\beta _+^2-\beta _-^2\right)^2}
\end{align}
and the solutions to the radial equation are:
\begin{align}
\label{hypersol}
    R_1(z)&=z^{\alpha } (1-z)^{\Delta /2} \, _2F_1\left[\frac{1}{2} (2 \alpha -2 \gamma +\Delta ),\frac{1}{2} (2 \alpha +2 \gamma +\Delta );1+2 \alpha;z\right]\notag \\
    R_2(z)&=z^{-\alpha } (1-z)^{\Delta /2} \, _2F_1\left[\frac{1}{2} (-2 \alpha -2 \gamma +\Delta ),\frac{1}{2} (-2 \alpha +2 \gamma +\Delta );1-2 \alpha ;z\right]
\end{align}
where $\Delta=1+ \sqrt{1+m^2}$. 
\subsection{Boundary conditions and Quantization}
The quantization condition at the boundary will depend on which of the above solutions we choose. In the case of a black hole, there are in-going and out-going boundary conditions that pick out one of the two solutions, and the quantization conditions can then be computed. However, in our case, there is no specific way of singling out one solution, and we can in general take the linear combination of both solutions, where the quantization condition will be dependent on both. In the case when the region containing CTCs is included, we argue that there is a natural boundary condition at $z=0$, where one chooses the solution that behaves regularly at this point. This boundary condition can be justified by matching the bulk result (\ref{moi}) with the dual CFT$_2$ result in the planar limit by analytically continuing the thermal two-point function of the planar BTZ black hole below the BTZ threshold with imaginary temperature. In the planar limit, the spinning defect geometry no longer contains CTCs in the region $r^2<0$, since $\phi$ coordinate is no longer periodically identified. Thus, we can make sense of this boundary condition without concerning ourselves with CTCs\footnote{I thank Hewei Frederic Jia for bringing this point to me.}.

As we shall see, the Feynman propagator will eventually sum over all possible quantized values of $\omega,\ell$. Since the two solutions only differ by a sign of $\alpha$, the final sum will take into account the contribution from both solutions. Thus, without loss of generality, we will choose $R_1(z)$ as our solution, and assume that $\alpha$ is positive in this case. 

On the other hand, if we choose to exclude the region containing CTCs, it is not clear what boundary condition we should impose at $r=0$, since $z$ will be regular at this point. As discussed above, we could take in general any linear combinations of the two solutions and solve for the spectrum accordingly. However, this procedure will not be tractable analytically. We will thus pick one of the solutions that would yield the same quantization condition as before and compute its normalization constant \footnote{It might still be possible that a very specific combination of the two solutions will yield the same two-point function as the one computed by including the CTCs. }. 

Thus, the asymptotic behavior of $R_1(z)$ near $z=1$ is:
\begin{align}
\label{asym}
    R_1(z)=& (1-z)^{-\Delta /2}\p{\frac{\pi  (z-1) \Gamma (1+2 \alpha ) \csc (\pi  \Delta )}{\Gamma (2-\Delta ) \Gamma \left(\alpha -\gamma +\frac{\Delta }{2}\right) \Gamma \left(\alpha +\gamma+\frac{\Delta }{2}\right)}+\mathcal{O}\p{(1-z)}}+ \notag\\[3pt]
    +&(1-z)^{\Delta /2}\p{\frac{\pi  \Gamma (1+2 \alpha ) \csc (\pi  \Delta )}{\Gamma (\Delta ) \Gamma \left(\alpha -\gamma -\frac{\Delta }{2}+1\right) \Gamma \left(\alpha +\gamma -\frac{\Delta }{2}+1\right)}+\mathcal{O}\p{{(1-z)}}}
\end{align}
In order for the function to have regular behavior at infinity, we impose the quantization condition\footnote{We will assume $\Delta > 2$. Other ranges of $\Delta$ lead to more general boundary conditions. See e.g. \cite{Marolf:2006nd} for a nice discussion of vector fields.}: 
\begin{align}
\label{quant}
    \frac{\Delta}{2}+\abs{\alpha} \pm \gamma=-k, \quad k \in \mathbb{Z}_+
\end{align}
The above quantization condition takes on different values depending on the range of parameter $M,J$
\begin{enumerate}
    \item $M<0, \abs{J}\leq \abs{M}$: $r_{\pm}$ are purely imaginary, and $\alpha,\gamma$ would be real:
    \begin{align}
        \omega_L&=\abs{\ell}\pm \p{2k+\Delta}\p{\beta_+-\beta_-} \notag\\
        \omega_R&=\abs{\ell}\pm \p{2k+\Delta}\p{\beta_+ +\beta_-} \notag
    \end{align}
    \item $M>0, \abs{J}\leq \abs{M}$: $r_{\pm}$ are real, and $\alpha,\gamma$ are purely imaginary. The above quantization condition represents the quasi-normal modes of the black hole:
    \begin{align}
        \omega_{L}&=\ell-i \left(r_+ - r_-\right) (\Delta +2 k)\notag \\
        \omega_{R}&=-\ell-i \left(r_-+r_+\right) (\Delta +2 k)\notag
    \end{align}
    which are associated with the in-going boundary condition.
    \item $M<0, \abs{J}>\abs{M}$: $r_{\pm}$ will have a purely imaginary and real part, and the values for $\alpha,\gamma$ are generally complex. In this case: 
    \begin{align}
        \omega_{L}&=\ell+ \sqrt{\frac{1}{n^2}+J}\hspace{2mm}(\Delta +2 k)\notag \\
        \omega_{R}&=-\ell-i \sqrt{J-\frac{1}{n^2}}\hspace{2mm} (\Delta +2 k)\notag
    \end{align}
    We see that $\omega_L$ is completely real but $\omega_R$ has an imaginary part. This perhaps suggests that these objects are not stable under perturbations. As discussed in \cite{Baake:2023gxx}, the quantum stress-energy tensor of a conformally coupled scalar field is non-renormalizable in this spacetime. 

\end{enumerate}

\subsection{Normalizability of the modes}
After obtaining the quantization condition, we can proceed to compute the normalization constant. We will simply quote the results and leave the details of the computation to Appendix \ref{Appen.A}. First, we restrict our attention to the region of the metric where $r^2>0$, and compute the normalization constant for the scalar fields. The integral is finite and can be obtained as a sum over Jacobi Polynomials of certain power evaluated at a specific value. However, it becomes difficult when one tries to perform the sum over $k$, which is a necessary step in constructing the propagator. On the other hand, if we extend the range of the radial coordinate into the region containing CTCs, i.e., $-\beta_-^2\leq r^2$, the normalization constant enjoys an algebraic simplification and can be obtained in a closed form as ratios of gamma functions, similar to the case with vanishing angular momentum. We will study the propagator including the region that contains CTCs, where the propagator can be expressed as an infinite sum over image geodesics, and the extrapolation of the leading divergence can be performed analytically. This divergence is insensitive to the region $r^2<0$ since the geodesic approximation that produces the same results does not probe the region of CTCs (see Appendix \ref{Appen.B}). However, the subleading divergences will be sensitive to the region containing CTCs, and we will provide further comments at the end of the next section.

The inner product defined on the constant time slice is given by:
\begin{align}
\label{normm}
\langle \Phi_I,\, \Phi_J \rangle &= -i \int_{\Sigma} \dd^2 x \sqrt{g_{\Sigma}(x)}\, n^{\mu} \Phi_I(x) \overset{\leftrightarrow}{\partial}_{\!\mu} \Phi_J^*(x) =\delta_{IJ}
\end{align}
where $I=(k,\ell)$, $\Sigma$ is a surface of constant $t$ slice, and $n^{\mu}$ is the future-pointing unit normal vector with respect to $\Sigma$. We will perform the integral in the transformed coordinate (\ref{cotrans}), where the metric is now given by: 
\begin{equation}
g_{\mu \nu}=\left(
\begin{array}{ccc}
 \frac{\beta _+^2-\beta _-^2 z}{z-1} & 0 & \beta _- \beta _+ \\
 0 & \frac{1}{4 (z-1)^2 z} & 0 \\
 \beta _- \beta _+ & 0 & \frac{\beta _-^2-\beta _+^2 z}{z-1} \\
\end{array}
\right)
\end{equation}
From the metric, we see that $\sqrt{g_{\Sigma}(x)}=\frac{1}{2} \sqrt{\frac{\beta _+^2 z-\beta _-^2}{(1-z)^3 z}}$ and the unit normal vector $n^{\mu}$ can be computed as:
\begin{align}
    n^{a}=\p{-\frac{g_{\phi \phi}}{g_{t \phi}}b\hspace{1mm},\hspace{1mm} 0\hspace{1mm},\hspace{1mm} b}, \hspace{6mm} b=\pm \p{-g_{tt}\p{\frac{g_{\phi \phi}}{g_t\phi}}^2+g_{\phi \phi}}^{-1/2}
\end{align}
We will choose the negative sign for $b$ so that the time component of the unit vector is positive. Now the normalization integral becomes:
\begin{align}
\label{norm2}
   \langle \Phi,\, \Phi \rangle &= -i \int \dd z \int_0^{2\pi} \dd \phi \sqrt{g_{\Sigma}(z)} \times n^{t} \hspace{1mm} \p{2i\omega} \hspace{1mm} \abs{R\p{z}}^2 \notag \\
    & \quad+i \int\dd z \int_0^{2\pi} \dd \phi \sqrt{g_{\Sigma}(z)} \times  n^{\phi} \hspace{1mm} \p{2i \ell} \hspace{1mm} \abs{R(z)}^2  \notag\\
    &=I_1+I_2
\end{align}
where we shall discard the constant factor $2\pi$ from the integral over $\phi$ by a redefinition of the normalization constant from now on. Note that the range of $z$ integration will depend on the regions of spacetime we wish to include. We will discuss the two cases separately below.

\subsubsection{$0\leq r^2 < \infty$}
In this region, the transformed radial coordinate has a range $c\leq z \leq 1$, where $c=\beta_-^2/\beta_+^2$, and the spacetime contains no CTCs. As discussed above, the Feynman propagator requires the sum over $\ell \in \mathbb{Z}$, and both branches of $\alpha$ will contribute to the propagator. Thus, we pick the $R_1(z)$ without loss of generality. As an illustration, we will evaluate the first integral $I_1$ and show that it is convergent at each value of $k$:
\begin{align}
\label{norm33}
  I_1 \propto  &\left. \sum_{s=2}^k \p{-1}^{s-1}\p{\partial_x^{k-s}g(x)}\partial_x^{s-1}\p{f(x)J_{k}^{\p{2\alpha,\Delta-1}}\p{x}} \right|^{1-2c}_{-1} + \notag \\
 & +(-1)^k \int_{-1}^{1-2c}\dd x g(x)\partial_x^k\p{f(x)J_{k}^{\p{2\alpha,\Delta-1}}\p{x}}
\end{align}
where we have $x=1-2z$, $f(x)=\frac{1-2c-x}{1-x}$, $g(x)=\p{1-x}^{2\alpha+k}\p{1+x}^{k+\Delta-1}$, and $J_{k}^{\p{2\alpha,\Delta-1}}(x)$ is the Jacobi Polynomial of $k$th order. The difficulty stems from the first term. For finite values of $k\in \mathbb{Z}_+$, this yields a finite sum of Jacobi polynomials of order $k-i$ evaluated at $x=1-2c$:
\begin{align}
  & \left. \sum_{s=2}^k \p{-1}^{s-1}\p{\partial_x^{k-s}g(x)}\partial_x^{s-1}\p{f(x)J_{k}^{\p{2\alpha,\Delta-1}}\p{x}} \right|_{1-2c}= \\
    =&\sum_{s=2}^k \sum_{i=0}^{s-1} \p{-1}^{s}\p{\partial_x^{k-s}g(x)}\frac{\Gamma\p{s}\Gamma\p{k+i+2\alpha+\Delta}}{2^{i}\Gamma\p{i+1}\Gamma\p{k+\alpha+\Delta}}\p{\frac{1}{\p{1-x}^{s-1-i}}}\left. J_{k-i}^{\p{2\alpha+i,\Delta-1+i}}\p{x} \right|_{x=1-2c} \notag
\end{align}
The expression is a finite sum over Jacobi polynomials and is finite at each value of $k$. This term comes from the integration by parts of the original integral, and cannot be simplified any further. When one performs the sum over $k$, the computation will not be tractable algebraically and the extrapolation of the leading divergence will be very difficult. The term involving the integral in (\ref{norm33}) can be evaluated in a closed form as ratios of gamma functions. As we shall see, when the region containing CTCs is included, the boundary term vanishes and it is precisely this integral that produces the normalization constant. The result of the second integral $I_2$ can be computed similarly, and the boundary term only differs by a sign compared to that of $I_1$ (See Appendix \ref{Appen.A}). It is tempting to expect that the boundary terms will cancel between the two integrals in (\ref{norm2}), but it is prevented by the prefactors in front of each integral.

\subsubsection{$-\beta_-^2\leq r^2 < \infty$}
In this region, the transformed radial coordinate has range $0\leq z\leq 1$, and the spacetime contains CTCs in the region $-\beta_-^2\leq r^2 <0$. We again choose $R_1(z)$ without loss of generality. The normalization constant can be computed exactly: 
\begin{align}
\label{normcons}
    \langle \Phi_I,\, \Phi_I \rangle \equiv& N_{k\ell}^2=\frac{\Gamma (2 \alpha +1)^2 \Gamma (k+1) \Gamma (k+\Delta )}{\Gamma (k+2 \alpha +1) \Gamma (k+2 \alpha +\Delta )}\mathcal{D}_{\ell k}, \notag \\[10pt]
    \mathcal{D}_{\ell k }=&\frac{\beta _+ \bigg(\ell \beta _- (\Delta +2 k)-\omega_{k\ell} \beta _+ \big(c (\Delta +2 \alpha +2k)-2 \alpha \big)\bigg)}{2 \alpha  \left(\beta _+^2-\beta _-^2\right) (\Delta +2\alpha +2k)}
\end{align}
where $\omega_{k \ell}$ is solved from the quantization condition (\ref{quant}). Without loss of generality, we choose: 
\begin{align}
\label{choosestuff}
    \alpha&=\frac{\ell \beta_+ -\omega \beta_-}{2(\beta_+^2-\beta_-^2)}\notag \\
    \omega_{k \ell}&=\ell+(2k+\Delta)\p{\beta_+-\beta_-} 
\end{align}
Together with $c=\beta_-^2/\beta_+^2$, the constant $\mathcal{D}_{\ell k}$ becomes:
\begin{align}
    \mathcal{D}_{\ell k}=\beta_+ +\beta_-
\end{align}
and with this simplification, we can move on to study the boundary propagator. 
\section{The boundary propagator}
\label{Section4}
The canonical quantization of the scalar fields proceeds as follows. The classical field is now viewed as an operator via the expansion:
\begin{align}
\label{eqn:full-solution}
\Phi(\mathbf{x}) = \sum_{\ell \in \Z} \sum_{k \in \N}\Big( a_{k\ell} e^{-i\w_{k\ell} t} e^{i\ell\phi} + a_{k\ell}^{\dagger} e^{i\w_{k\ell} t} e^{-i\ell\phi} \Big) R(z), \quad \big[a_{k\ell},\, a_{k'\ell'}^{\dag}\big] = \d_{kk'} \d_{\ell\ell'}.
\end{align}
Here we choose $\omega_{k\ell}>0$. This choice determines the raising and lowering operator of the theory. The bulk-to-bulk Wightman function can then be assembled as:
\begin{align}
\label{eqn:wightman0}
G^+(\mathbf{x},\mathbf{x}') = \mel{0}{\Phi(\mathbf{x}) \Phi(\mathbf{x}')}{0} = \sum_I \Phi_I(\mathbf{x}) \Phi_I^*(\mathbf{x}'), \qquad G^-(\mathbf{x},\mathbf{x}') = G^+(\mathbf{x}',\mathbf{x}).
\end{align}
where the summation index $I=\p{k,\ell}$. The Feynman propagator $G(t,\th)$ is constructed from the Wightman functions $G^{\pm}(t,\phi)$:
\begin{equation}
\label{eqn:heaviside}
\begin{aligned}
G(t,\phi) = \Theta(t) G^+(t,\phi) + \Theta(-t) G^+(-t,\phi).
\end{aligned}
\end{equation}
The boundary-to-boundary propagator can then be obtained from the extrapolate dictionary \cite{Witten:1998qj,Gubser:1998bc}, and we normalize each of the $\Phi$ modes with a factor of $N_{k\ell}^{-1}$ computed above: 
\begin{align}
    \label{eqn:wightman1}
G^+(t,\phi) = \lim_{z\to 1} \! \Big( \p{1-z_1}^{\Delta/2}\p{1-z_2}^{\Delta/2}  G^+(z_1,z_2) \Big) = \sum_{\ell \in \Z} \sum_{k \in \N} e^{-i\w_{k\ell} t} e^{i\ell\phi} \Big( N_{k\ell}^{-1} A^-_{k\ell}(\Delta) \Big)^2.
\end{align}
where from (\ref{asym}) we have: 
\begin{align}
    A^-_{k\ell}(\Delta)=\frac{\pi ^2 \Gamma (2 \alpha +1)^2 \csc ^2(\pi  \Delta )}{\Gamma (\Delta )^2 \Gamma (k+2 \alpha +1)^2 \Gamma (1-k-\Delta )^2}
\end{align}
Combining everything: 
\begin{align}
\label{finalpropagator}
    G^+(t,\phi)&=\sum_{\ell \in \Z} \sum_{k \in \N} e^{-i\w_{k\ell} t} e^{i\ell\phi} C_{k\ell}^2, \notag \\
    C_{k\ell}^2=\Big( N_{k\ell}^{-1} A^-_{k\ell}(\Delta) \Big)^2 &=\frac{\Gamma\p{\Delta+k+2\alpha}\Gamma\p{\Delta+k}}{\mathcal{D}_{k\ell}\Gamma\p{\Delta}^2\Gamma\p{1+k+2\alpha}\Gamma\p{1+k}}
\end{align}
We show in Appendix.\ref{Appen.C} using spectral decomposition that the above expression is equivalent to: 
\begin{align}
\label{moi}
    G^{+}\p{t,\phi}=\sum_{m\in \mathbb{Z}} \bigg[ \frac{4}{\beta_+^2-\beta_-^2}\sin\p{\frac{\beta_++\beta_-}{2}\p{\phi+2\pi m-t}}\sin\p{\frac{\beta_+-\beta_-}{2}\p{\phi+2\pi m+t}} \bigg]^{-\Delta}
\end{align}
up to appropriate renormalization due to the infinite sum of the image geodesics. Several comments are now in order. 
\begin{enumerate}
\item It is worth emphasizing that although one could easily obtain the expression (\ref{moi}) using the method of images, it is not obvious that the geometry that yields this expression contains CTCs, which originates from the range of the radial integration in the normalization constant. This can be more easily understood from the perspective of the covering space. As discussed in \cite{Banados:1992gq}, it is quite unnatural to exclude the region where $\Theta \cdot \Theta \leq 0$ in the covering space. However, since the surface $\Theta \cdot \Theta =0 \hspace{2mm} (r^2=0)$ appears as the singularity of the BH/Defect spacetime, one could instead view this surface as the actual singularity of the causal structure of the spacetime, despite the fact that the entire manifold is everywhere negatively curved. Similarly in our case, excluding the region containing CTCs corresponds to imposing an artificial cutoff in the global AdS$_3$ spacetime (\ref{coortrans}), and the boundary terms of the normalization integral will no longer vanish at this cutoff. This essentially prevents us from writing the propagator as a sum over image geodesics. 

\item In the case of vanishing angular momentum, we can express the propagator as a finite sum over image geodesics if the angle deficit parameter $n$ is a positive integer \cite{Berenstein:2022ico}. However, in the case of a spinning defect, by writing it as an infinite sum over image geodesics, we do not need to impose any conditions on the parameters $\beta_+$ or $\beta_-$, since the infinite series guarantees the periodicity of the propagator. This generalizes trivially to the case of vanishing angular momentum where by writing the propagator as an infinite sum over image geodesics, the angle deficit parameter $n$ is no longer constrained to integer values (See Appendix.\ref{Appen.C} for more details). 

\item The necessity of regularizing the sum should not be surprising. In the dual CFT$_2$, such sum over infinite image geodesics is interpreted as summing over vacuum blocks across all channels. This is in the same spirit as the Poincare/Farey Tail sum, which in many cases will diverge, and must be regulated \cite{Maloney:2016kee}. Thus, we are simply extrapolating the finite piece, which is the mode sum in the bulk. Schematically speaking, when applying the spectral decomposition to the infinite sum over $m$, one essentially obtains a Dirac-Delta comb that selects out the quantization condition: 
\begin{align}
    \sum_{\ell,k}\sum_{p}\delta\p{\text{0}}_p \times \p{\text{Mode Sum}}_{\ell,k}
\end{align}
where the Mode Sum represents the ratios of gamma functions. When the quantization condition is satisfied, the delta function is evaluated at 0 and becomes infinite, and we can simply divide out this infinity term by term in $p$. In contrast, when the sum $m$ is finite, as in the case for the non-spinning defect with integer angle-deficit, the Dirac-Delta comb becomes the Kronecker comb, and the infinity is no longer present. One can similarly apply certain conditions to $\beta_+,\beta_-$, so that the sum over image geodesics becomes finite (See Appendix.\ref{Appen.C} for more details).
\end{enumerate}
The leading singular behavior can easily be seen as: 
\begin{align}
\label{propagator1}
    G^+(t,\phi) \approx  \bigg[\frac{2}{\beta_+^2-\beta_-^2} \Big(\! \cos\p{t\beta_+- \phi \beta_-} - \cos\p{t\beta_- -\phi\beta_+}\Big)\bigg]^{-\Delta}.
\end{align}
This matches the geodesic approximation computation for the propagator in Appendix.\ref{Appen.B} as expected, since the geodesic computation can be trusted when the mass of the scalar fields becomes large $\Delta \sim m$. It is also instructive to compute the short-distance behavior of the propagator at equal times, when the two insertion points on the boundary are brought close to each other. In the bulk, this corresponds to the two scalar fields having a large relative angular momentum $\ell$. By taking $\ell \rightarrow \infty$ in (\ref{finalpropagator}), performing the sum over $k,\ell$, and then expanding again in the limit $\phi \rightarrow 0$:
\begin{align}
\label{expansions}
    G(\phi)\sim \left(\phi \right)^{-2\Delta } \left(1+\frac{\Delta  \left(\beta _-^2+\beta _+^2\right) \phi ^2}{{12 }}+  \right. \hspace{6cm} \\ \notag 
     + \left. \frac{\Delta  \left(\beta _-^4+6 \beta _+^2 \beta _-^2+\beta _+^4+5 \Delta 
   \left(\beta _-^2+\beta _+^2\right){}^2\right) \phi ^4}{1440 }+O\left(\phi ^6\right)\right)
\end{align}
where the leading divergences can then be compared term by term with the field theory computation. When the propagator is written in terms of sum over images, the above leading divergence is completely captured by the minimal geodesic as expected. 

\section{The CFT interpretation}
\label{section5}
The scalar field propagator has a nice CFT interpretation in terms of semi-classical heavy-light correlator \cite{Fitzpatrick:2014vua,Fitzpatrick:2015zha}. The propagator is recast as a four-point function of the spinning particle and scalar fields. The coefficient of the mode sum we found before takes the form of the s-channel OPE, which is dominated by the exchange of the “double-trace” primaries $\mathcal{O}_{k,\ell}=\left[\mathcal{O}_H\mathcal{O}_L\right]_{k,\ell}$. These states are precisely created by the scalar field orbiting around the spinning particle, where in the limit of large separation, the interactions (binding energy) between these two objects are negligible, and the back-reaction of the light scalar field on the geometry can be ignored. The field theory essentially reduces to that of a  generalized free theory (GFT), which our bulk computation is dual to. The four-point function can also be computed in the t-channel, where the in the large $c$ limit, assuming that $h_H, \bar{h}_H\propto c \gg h_L, \bar{h}_L$, the Virasoro conformal block was shown to be equivalent to the global conformal block \cite{Fitzpatrick:2015zha}, and it captures the descendants of the stress-energy tensor and represents the exchange between boundary graviton with the spinning particle and the scalar fields. We will first review the equivalence of the mode sum computation in the bulk with the s-channel expansion. Then, we will move on to discuss the t-channel computation, where we apply the results of \cite{Fitzpatrick:2015zha} to the spinning particle state, thus verifying the crossing symmetry. The subleading divergences can also be matched with the bulk computation by summing over the modular $PSL\p{2,\mathbb{Z}}$ contributions to the conformal block, which corresponds to the method of images in the bulk in the semi-classical limit \cite{Maloney:2016kee}. 
\subsection{HHLL Four-Point function}
We will work in the radial quantization, and adopt the complex coordinates $z=e^{\tau+i \phi}$ and $\bar{z}=e^{\tau-i \phi}$, where $i\tau=t$. The spinning particle is dual to the heavy operator with conformal dimension and spin given by:
\begin{align}
\label{confordandspin}
    h_H+\bar{h}_H=\frac{c}{12}\p{1-\frac{1}{n^2}},\quad h_H-\bar{h}_H=\frac{c}{12}J
\end{align}
which are the ADM mass and angular momentum normalized with respect to the AdS vacuum respectively. For convenience in our later discussion, we define: 
\begin{align}
\label{alphadef}
    \alpha_H=\sqrt{1-\frac{24h_H}{c}}, \quad \bar{\alpha}_H=\sqrt{1-\frac{24\bar{h}_H}{c}}
\end{align}
From the operator-state correspondence, the heavy operator $\mathcal{O}_H$ defines a state via $|\mathcal{O}_H\rangle=\mathcal{O}_H|0\rangle$, which is dual to the spinning defect geometry. We would like to make two insertions of the heavy operator at $z=0,z=\infty$ that correspond to the defect existing for all times. The scalar propagator computed in the background of the heavy spinning particle can then be recast as a normalized four-point function:
\begin{align}
\label{eqn:correlator-def}
G(z, \bar{z}) = 
\frac{\big\langle \mathcal{O}_H(\infty) \mathcal{O}_L(1) \mathcal{O}_L(z, \bar{z}) \mathcal{O}_H(0)\big\rangle }{\big\langle \mathcal{O}_H(\infty) \mathcal{O}_H(0)\big\rangle}=\big\langle \mathcal{O}_H(\infty) \mathcal{O}_L(1) \mathcal{O}_L(z, \bar{z}) \mathcal{O}_H(0)\big\rangle_{\mathcal{N}}
\end{align}
This correlator can be evaluated through either the s or the t-channel, each summing up the Virasoro conformal blocks associated with all the local primary operators within the theory. In the s-channel, the scalar field $\Phi$ has a low conformal dimension compared with the heavy operator in the large $c$ limit, and the s-channel is dominated by the double-trace primaries $\mathcal{O}_{k,\ell}=\left[\mathcal{O}_H\mathcal{O}_L\right]_{k,\ell}$ with dimension \cite{Fitzpatrick:2014vua}: 
\begin{align}
\label{doubletrace}
    h_{k\ell}\approx h_H+\alpha_H\p{h_L+k}+\ell, \quad \bar{h}_{k\ell}\approx h_H+\bar{\alpha}_H\p{h_L+k}+\ell
\end{align}
where we have neglected the terms suppressed by $1/c$. In other words, one expects that the OPE to be dominated by primaries operators of the form $\mathcal{O}_H \partial^{[k]}\mathcal{O}_L$, where the recoil effect of $O(1/c)$ are suppressed in the large $c$ limit \cite{Berenstein:2021pxv} (See also \cite{Berenstein:2014cia,Berenstein:2019tcs}), which is an equivalent way of stating that the back-reaction of the light scalar field on the geometry is negligible. The conformal dimension of the double-trace operator is precisely $\Delta_H$ plus the quantization condition (\ref{quant}) found in the bulk computation, which captures the energy excitation of the scalar field orbiting around the heavy particle with excitation number $k$ and angular momentum $\ell$. Furthermore, the OPE coefficient is related to the normalization constant computed in the bulk. Thus, the leading expansion in the s-channel can be shown to be equivalent to the bulk mode sum. 

\subsection{s-channel Expansion}
The equivalence between the s-channel expansion and the bulk mode sum was already discussed in great detail in \cite{Berenstein:2022ico}. Here we briefly review the discussion. The s-channel expansion of the four-point function can be written as: 
\begin{align}
   G(z,\bar{z})=\sum_{h,\bar{h}} \big\langle \mathcal{O}_H(\infty) \mathcal{O}_L(1) P_{h,\bar{h}} \mathcal{O}_L(z, \bar{z}) \mathcal{O}_H(0)\big\rangle_{\mathcal{N}}=\sum_{h,\bar{h}} \mathcal{C}^2_{HLh} \mathcal{F}_{h}\p{z} \tilde{\mathcal{F}}_{\bar{h}}\p{\bar{z}}
\end{align}
where the identity operator is given by:
\begin{align}
    P_{h,\bar{h}}=\sum_{\alpha_{h,\bar{h}}} |\alpha_{h,\bar{h}} \rangle \langle \alpha_{h,\bar{h}} |
\end{align}
Here we have organized the states into irreducible representations of the Virasoro group, where the states $| \alpha_{h,\bar{h}}\rangle$ are the primary operators associated with the irreducible representations as well as their descendants. The basic idea is that the s-channel four-point block has a simple singular behavior at $z=0$: 
\begin{align}
    \mathcal{F}_{h}\p{z}=z^{h-h_H-h_L}\p{1+\sum_{n=1}^{\infty} c_nz^n}
\end{align}
for some coefficient $c_n$. The bulk computation corresponds to primary exchange, where we simply approximate it with the leading behavior. If we choose $h=h_{k\ell}$ (\ref{doubletrace}), the intermediate states precisely correspond to the double-trace primary exchange, and the four-point function can then be approximated by:
\begin{align}
     G(z,\bar{z}) &\approx \sum_{h,\bar{h}}\mathcal{C}_{HLh}^2 z^h \bar{z}^{\bar{h}} \abs{z}^{-\Delta_H - \Delta_L} \notag \\
    &=\sum_{h_{k\ell}}\mathcal{C}_{HLh}^2 e^{-i\omega_{k\ell}}e^{i\ell \phi}
\end{align}
where we have transformed back to ($t,\phi$) coordinate. If we choose $\mathcal{C}_{HLh}=C_{k\ell}$, this precisely corresponds to the bulk mode sum (\ref{finalpropagator}). Note that the OPE coefficient $\mathcal{C}_{HLh}$ can be viewed as the normalization of the three-point function $\langle \mathcal{O}_H \mathcal{O}_L \mathcal{O}_h \rangle$, which is by definition dependent on the bulk geometry, and its form will be dependent on whether we choose to include the region containing CTCs or not.

\subsection{t-channel Expansion}
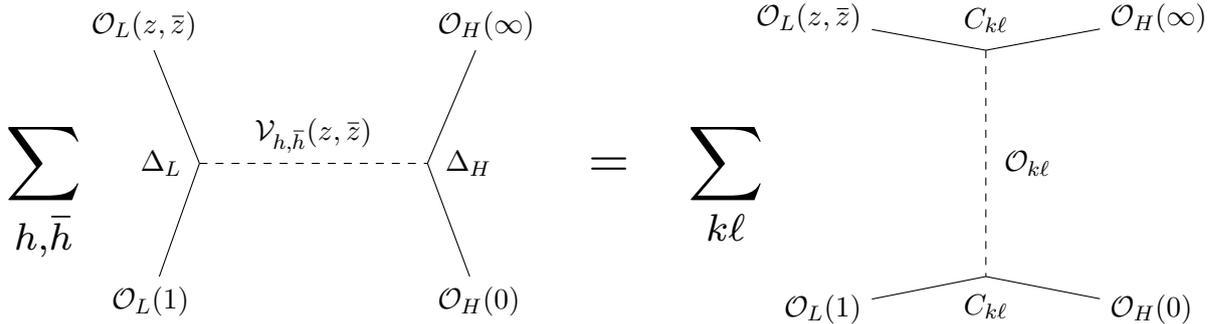
\begin{figure}[t]
\centering
\[
\scalebox{1.8}{$\displaystyle  \sum_{h,\bar{h}} $}
\begin{gathered}
\begin{tikzpicture}
    \begin{feynman}
    \vertex (a);
    \vertex [left = 0.1cm of a] (a1) {\( \Delta_L \)};
    \vertex [above left = 1.5cm and -0.1cm of a] (b) {\(\mathcal{O}_L(z, \bar{z})\)};
    \vertex [below  left = 1.5cm and -0cm of a] (d) {\(\mathcal{O}_L(1)\)};
    \vertex [right = 3cm of a] (f);
    \vertex [right = 0.1cm of f] (g) {\( \Delta_H \)};
    \vertex [above right = 1.5cm and -0cm of f] (c) {\(\mathcal{O}_H(\infty)\)};
    \vertex [below right = 1.5cm and -0cm of f] (e) {\(\mathcal{O}_H(0)\)};
    \diagram* {
        (b) -- (a) -- (d),
        (a) -- [scalar, edge label = {$\mathcal{V}_{h,\bar{h}}(z, \bar{z})$}
        ] (f),
        (c) -- [plain] (f) -- [plain] (e),
    };
    \end{feynman}
\end{tikzpicture}
\end{gathered}
\scalebox{1.8}{$\displaystyle  \;\;\; = \;\;\; \sum_{k\ell}\!\!$}
\begin{gathered}
\begin{tikzpicture}
    \begin{feynman}
    \normalsize
    \vertex (a);
    \vertex [above left = 0.1cm and 1.5cm of a] (a1) {\( \mathcal{O}_L(z, \bar{z}) \)};
    \vertex [above right = 0.1cm and 1.5cm of a] (a2) {\( \mathcal{O}_H(\infty) \)};
    \vertex [above = 0.1cm of a] (c1) {$C_{k\ell}$};
    \vertex [below = 3cm of a] (b);
    \vertex [below left = 0.1cm and 1.5cm of b] (b1) {\( \mathcal{O}_L(1) \)};
    \vertex [below right = 0.1cm and 1.5cm of b] (b2) {\( \mathcal{O}_H(0) \)};
    \vertex [below = 0.1cm of b] (c2){$C_{k\ell}$};
    \diagram* {
    (a1) -- (a) -- (a2),
    (a) -- [scalar, edge label = $\;\mathcal{O}_{k\ell}$] (b),
    (b1) -- (b) -- (b2),
    };
    \end{feynman}
\end{tikzpicture}
\end{gathered}
\]
\vspace{-1pt}
\caption{These diagrams illustrate the crossing equation and the kinematics of our OPEs. The t channel (left) is dominated by the vacuum block with $h=\bar{h}=0$, while the s channel (right) is dominated by the double-trace primaries. On the right, we have labeled the OPE coefficients that correspond to the 3-point functions $\big\langle \mathcal{O}_H \mathcal{O}_L \mathcal{O}_{k\ell} \big\rangle$. Figure taken from \cite{Berenstein:2022ico}.}
\label{fig:feynman-diagram}
\end{figure}

We can similarly expand in the t-channel: 
\begin{align}
    G(z,\bar{z})=\sum_{h,\bar{h}} \big\langle \mathcal{O}_H(\infty) \mathcal{O}_H(0) P_{h,\bar{h}} \mathcal{O}_L(z, \bar{z}) \mathcal{O}_L(1)\big\rangle_{\mathcal{N}}=\sum_{h,\bar{h}}\mathcal{V}_{h}\p{z}\bar{\mathcal{V}}_{\bar{h}}\p{\bar{z}}
\end{align}
A powerful method for computing the Virasoro conformal block in the t-channel was proposed in \cite{Fitzpatrick:2015zha}, where it was shown that the complete Virasoro conformal block for a semiclassical heavy-light correlator is equivalent to the global conformal block \cite{Dolan:2000ut,Dolan:2003hv,Dolan:2011dv}, where the light operators are assessed in a different set of coordinates: $\omega=z^{\alpha_H}$, such that $\mathcal{O}_L(z) = (\omega'(z))^{h_L} \mathcal{O}_L (w(z))$, and the conformal block reads:
\begin{align}
    \mathcal{V}_{h}(z)=z^{\p{\alpha_H-1} h_L}\p{\frac{1-z^{\alpha_H}}{\alpha_H}}^{h-2h_L} {}_2 F_1 \Big[h,\, h,\, 2h;\, 1-z^{\alpha_H}\Big]
\end{align}
where $\alpha_H$ is given in (\ref{alphadef}) and similarly we have the anti-holomorphic piece $\bar{\mathcal{V}}_{\bar{h}}(\bar{z})$. The basic idea is that after the coordinate transformation, the light scalar field is now evaluated on a thermal background created by the heavy operator, which changes the motion of the light fields around it. This is very similar to the bulk side, where the monodromy of $z^{\alpha_H}$ determines the geometry in the bulk. The conformal block in the t-channel is dominated by the identity operator $h,\bar{h}=0$ in the regime of our interests, and we can approximate the normalized four-point function using the vacuum block $\mathcal{V}_{0}(z)$: 
\begin{align}
\label{four-point}
    \big\langle \mathcal{O}_H(\infty) \mathcal{O}_L(1) \mathcal{O}_L(z, \bar{z}) \mathcal{O}_H(0)\big\rangle_{\mathcal{N}} \approx \mathcal{V}_{0}(z) \bar{\mathcal{V}}_{0}(\bar{z}), \quad \mathcal{V}_{0}(z)=z^{\p{\alpha_H-1} h_L}\p{\frac{1-z^{\alpha_H}}{\alpha_H}}^{-2h_{L}}
\end{align}
where in the case of a spinning particle, we have: 
\begin{align}
    \alpha_H=(\beta_++\beta_-), \quad \bar{\alpha}_H=\p{\beta_+-\beta_-}
\end{align}
The first consistency check we can make is to consider the UV behavior of the four-point function by studying the expansion when the two light scalars are brought close to each other at equal times. Transforming back to the real-time coordinate $(t,\phi)$, the expansion matches exactly to (\ref{expansions}). Moreover, at different times, (\ref{four-point}) exactly reproduces the leading divergence of the bulk propagator (\ref{propagator1})\footnote{There is a factor of $z^{-\Delta}$ difference originating from the primary fields transforming from the plane to the cylinder.}: 
\begin{align}
     \big\langle \mathcal{O}_H(\infty) \mathcal{O}_L(1) \mathcal{O}_L(z, \bar{z}) \mathcal{O}_H(0)\big\rangle_{\mathcal{N}}= \bigg[\frac{2}{\beta_+^2-\beta_-^2} \Big(\! \cos\p{t\beta_+- \phi \beta_-} - \cos\p{t\beta_- -\phi\beta_+}\Big)\bigg]^{-\Delta}
\end{align}
Since the sum over modes in the bulk is equivalent to the s-channel computation, this provides a check of the crossing equation (See Figure.\ref{fig:feynman-diagram}). This matching highlights the connection between the Virasoro blocks and bulk geodesics, which was extensively investigated in \cite{Asplund:2014coa,Hijano:2015zsa,Hijano:2015qja}, where the emergence of bulk Witten diagrams from the Virasoro blocks is elucidated.

\subsection{Subleading divergences}
The matching of the leading divergence in the t-channel should not be surprising, since this term can be more easily understood as an analytical continuation from the thermal two-point function with imaginary temperature, which is the exact answer in the planar limit when we decompactify $\phi$: $-\infty<\phi<\infty$. The region $r^2<0$ no longer contains CTCs in the planar limit, and the matching of our bulk computation with the analytical continuation of the thermal two-point function justifies our choice of boundary condition imposed at $z=0$ in \S.\ref{section3}. The thermal two-point function of CFT$_2$ on a line can be obtained from simple coordinate transformation independent of the knowledge of the bulk. However, when $\phi$ is periodic, there will be subleading contributions that cannot be obtained from first principles from the dual CFT$_2$, and the bulk computation provides the definition for the two-point function by the method of images.

The language of the HHLL four-point function will become important when the subleading divergences are taken into account. As discussed in \cite{Hijano:2019qmi}, one would naively expect that the propagator of two light primary operators in the bulk is computed solely from the vacuum block between the light and heavy primary operators. However, this is not correct due to the lack of periodicity or more precisely the fact that the vacuum block is not single-valued as we move the light field around the heavy defect. The bulk propagator restores the periodicity in $\phi$ by the method of images, and one natural question to ask is whether the same is also true in the dual field theory. This direction has been explored in \cite{Maloney:2016kee}, and the basic idea is to exploit the modular structure of the conformal blocks. Instead of writing the conformal blocks as a function of the cross-ratio $x$, it was proposed to view the cross-ratio $x$ as an image of a point $\tau$ in the upper-half plane $\mathbb{H}_+$. Since the map between $x$ and $\tau$ does not have a unique inverse map $\tau\p{x}$, the space of $x$ is then viewed as a quotient of $\mathbb{H}_+$ by some group, and the crossing symmetry can be viewed as modular invariance of $\mathbb{H}_+$. In order to have a crossing symmetric four-point function in the semi-classical limit, one should then average the contribution from the vacuum block over the modular group $PSL\p{2,\mathbb{Z}}$. This can be viewed as taking into account the vacuum
blocks across all channels and summing over all of them. Schematically, the four-point function is given by \cite{Maloney:2016kee}: 
\begin{align}
    \big\langle \mathcal{O}_H \mathcal{O}_L \mathcal{O}_L \mathcal{O}_H\big\rangle \sim \sum_{\substack{\text{classical} \\ \text{solutions}}}\sum_{\substack{\text{light} \\ \text{primaries}}} \mathcal{F}_h
\end{align}
where the sum over classical solutions can be viewed as the sum over saddle points of the bulk. This is very similar to the Farey tail/Poincare sum in the computation of the partition function in 3D gravity \cite{Maloney:2007ud,Keller:2014xba,Dijkgraaf:2000fq,Manschot:2007ha}. The saddle points in our case are precisely the non-minimal geodesics in the bulk, and the entire bulk physics is captured by the vacuum block across all channels. Thus, following the proposal \cite{Maloney:2016kee}, we have: 
\begin{align}
    G(z,\bar{z})=\sum_{m\in \mathbb{Z}}\mathcal{V}_{0}\p{z \cdot e^{2\pi i m \alpha_H }}\bar{\mathcal{V}}_{0}\p{\bar{z}\cdot e^{-2\pi i m \bar{\alpha}_H}}
\end{align}
This precisely reproduces the sum over images in the bulk propagator (\ref{moi}), which would suggest that the bulk geometry corresponding to state (\ref{confordandspin}) will include the region containing CTCs, since the normalization constant that produces this result is computed including region with CTCs. As we have seen in the bulk, excluding the region of CTCs in the quotient space is quite unnatural from the perspective of the covering space, which makes it difficult to write the propagator using the method of images due to the complicated expression of the boundary terms, and it is tempting to conjecture that geometries dual to this state will contain CTCs due to the uniqueness of the BTZ metric. However, it might still be possible that the method of images can be extrapolated from the complicated expression (\ref{norm33}) or perhaps using a different quantization procedure. 
\section{Summary and Discussion}
\label{section6}
In this note, we studied the spinning particle/defect geometry in the context of AdS$_3$/CFT$_2$. We solved the bulk equation of motion, computed the quantization condition, and constructed the propagator. We demonstrated that by including the region containing CTCs, there is a natural boundary condition to impose in the bulk, and the normalization constant of the scalar field can be obtained analytically. The propagator can then be written as an infinite sum over image geodesics up to an appropriate normalization. From the dual CFT$_2$, we reviewed that the s-channel computation is equivalent to the mode sum in the bulk, and that the contribution from the sum of the t-channel vacuum block across all channels reproduces the bulk propagator that was computed including the region containing CTCs. Nevertheless, certain aspects of our work require further clarification.

\subsection*{Bulk causality} We have demonstrated that the normalization constant as well as the propagator become algebraically manageable when the region containing CTCs is included, and the propagator can be written as a sum over image geodesics. This simplification stems from the identification procedure to obtain the spinning particle geometry from pure AdS$_3$. As is clear from (\ref{coortrans}), if we choose to include the CTCs region in the spinning geometry, this would correspond to the more natural range of radial coordinate in the covering space, where we have $0\leq \hat{r}<\infty$, and the boundary terms vanish at $\hat{r}=0$. Therefore, we have a natural boundary condition to impose at $z=0$. However, by excluding the CTCs region in the spinning geometry, there is no simple boundary condition one can impose at $r=0$ that singles out one of the solutions, and we can in general take the linear combination of both solutions where analytical computation will be more intractable. However, if we choose to pick the solution that yields the same quantization as before, we are essentially imposing an artificial cutoff for the radial coordinate in the covering space, where the boundary terms no longer vanish at this cutoff, as we have seen in \S.\ref{section3}. It is not clear to us if the region containing CTCs bears any other deeper physical significance, but a more thorough understanding of these regions will certainly provide valuable insights into the causal structure of the bulk geometry that is dual to the state (\ref{confordandspin}).

\subsection*{Subleading contribution} Recent investigations, such as those presented in \cite{Grabovsky:2024jwf} for a heavy defect with zero angular momentum, have demonstrated that the OPE coefficient of the heavy-heavy/light-light (HH/LL) double-trace operator can be matched to the contributions from non-minimal geodesics in the bulk propagator. The approach presented in \cite{Grabovsky:2024jwf} is certainly more in tune with the conventional CFT viewpoint, where one sums over operators in a fixed channel. On the other hand, \cite{Maloney:2016kee} suggests that similar to the partition function, the correlation function should be computed by summing across channels. This is justified if the vacuum blocks across different channels do not overlap when dualized in a single fixed channel \cite{Anous:2017tza}. Thus, it would be interesting to understand the relationship between these two approaches. Moreover, extending the analysis in \cite{Grabovsky:2024jwf} to scenarios with non-zero angular momentum would be interesting. However, similar to the s-channel expansion, the computations of the OPE coefficients of the HH/LL double trace operator rely on results obtained in the bulk, and will not provide further insights into the region containing CTCs.

\subsection*{Extremal/Overspinning particles} It would also be interesting to understand the extremal spinning particle. These solutions admit a globally defined Killing spinor, and can be interpreted as extremal 0-branes which saturate the BPS bound \cite{Miskovic:2009uz}, They might play an important role in the supersymmetric extension of (2+1) dimensional gravity. Additionally, the BTZ family of metrics also admits overspinning solutions, which is casually well-behaved classically in the region $r^2>0$. However, these solutions are more problematic when quantum effects are introduced \cite{Baake:2023gxx}, and might not have an interpretation in the dual unitary CFT$_2$ due to its overspinning nature \cite{Basile:2023ycy}. 

\subsection*{BH creation} Another aspect worth investigating is whether the non-extremal spinning solutions can create black holes. Classically, the non-spinning defect can produce rotating black holes when they have a non-zero impact parameter \cite{Holst:1999tc}, and the final geometry is represented by the BTZ metric. It would be interesting to see if the non-extremal spinning solutions could also produce black holes from collisions with other defects when the total energy is above a certain threshold. Intuitively, this process could be allowed if the mass of the defects alters the singularity's monodromy from elliptic to hyperbolic, thereby facilitating the formation of a black hole. Moreover, it would be interesting to obtain a description of the BH creation process from the dual field theory. Tentatively, the collision of two heavy defects forming a black hole might be related to the heavy-heavy-heavy three-point function recently studied in \cite{Abajian:2023jye,Abajian:2023bqv}, where one has three insertions at $z=0,1,\infty$. This can be viewed in the bulk as having a defect in the past $t=-\infty$ ($z=0$), inserting another defect at $t=0$ ($z=1$), and these two form a BH in the future $t=\infty$ ($z=\infty$). Since the weights of the operators at these insertion points are arbitrary, one can match the monodromy of these singularities to that of the bulk, thus establishing a relation between the masses of the defects and the final mass of the BH, and obtaining Gott's condition. The monodromy matching was done in \cite{Welling:1995er}, where Gott's condition was obtained via the matching procedure.

\subsection*{Flat-space limit}
The study of the flat space limit of AdS/CFT has a long and interesting history. It involves transforming the AdS amplitude in the flat-space limit, which enables the extraction of the flat-space S-matrix for scattering processes \cite{Polchinski:1999ry,Giddings:1999jq,Gary:2009mi,Penedones:2010ue} \footnote{The idea goes back to Dirac \cite{Dirac:1936fq} and was rediscovered by Weinberg \cite{Weinberg:2010fx}.}. Although quantum field theory (QFT) in AdS with a large AdS radius essentially reduces to QFT in flat-space, taking such a limit in terms of correlation functions is a non-trivial task. In particular, \cite{Hijano:2019qmi} studied the flat space limit of the correlator in a CFT$_2$ deﬁcit state (See \cite{Duary:2023gqg} for a recent study of flat-space limit of correlators in AdS$_2$). It was found that using a specific prescription of taking the flat space limit for the CFT$_2$ correlator, one essentially obtains the flat-space S-matrix scattering on a cone.  Investigating whether this approach can be extended to cases with angular momentum is interesting, especially given the known metric for a spinning particle in flat spacetime \cite{Deser:1983tn}.  

\subsection*{Entanglement Entropy}
Twist operator correlation function (TOC) plays an important role in the studies of entanglement entropy \cite{Lunin:2000yv,Calabrese:2004eu}, and more recently it has been associated with a canonical algebro-geometric object, the isomonodromic tau function, which is achieved using a generalized stress-tensor method of Calabrese and Cardy, thus bestowing the TOC with an integrable system interpretation \cite{Jia:2023ais} (See also \cite{Jia:2023dpt}). In \cite{Berenstein:2022ico}, utilizing the fact that the weight of the twist operator becomes light compared with that of the heavy defect in the limit $n \rightarrow 1$, it was argued that one can essentially approximate the TOC by the free field computation in the bulk. However, this does not yield the desired R\'enyi entropy, and one encounters the order of limit issue in the $n \rightarrow 1$ limit \cite{Fischetti:2014zja}. It would thus be interesting to understand these subtleties and explore the multi-interval R\'enyi/entanglement entropy in the spinning state, and obtain the corrections near the phase transition along the lines of \cite{Marolf:2020vsi,Dong:2020iod,Akers:2020pmf}.

\acknowledgments
I am grateful to David Berenstein for many insightful discussions, his patience, and his encouragement. I would also like to thank Steven Carlip, Gary Horowitz, Jesse Held, Adolfo Holguin, Maciej Kolanowski, Sean McBride, and  Zhencheng Wang for helpful discussions. I owe a special thanks to David Grabovsky and Hewei Frederic Jia for many thorough discussions and for explaining their papers \cite{Grabovsky:2024jwf}, \cite{Jia:2023ais,Jia:2023dpt} to me. It is a pleasure to thank David Berenstein and Adolfo Holguin for detailed comments and feedback on a draft of the paper. I wish to express my gratitude for the hospitality and inclusiveness of the Department of Physics at UCSB, where this work was completed. This work is supported in part by funds from the University of California.

\appendix

\section{Computation of integrals}
\label{Appen.A}
We compute the normalization integrals in \S.\ref{section3}. 
\subsection{$0\leq r^2 < \infty$}
We pick the $R_1(z)$ as our solution, and use the quantization condition $\frac{\Delta}{2}+\alpha-\gamma=-k$, with $\gamma=\p{\omega \beta_+-\ell \beta_-}/\p{2\p{\beta_+^2-\beta_-^2}}$ which corresponds to the choice $\omega_{\ell k}>0$ consistent with the case of vanishing angular momentum. Plugging this qunatization condition into our solution, the integral becomes: 
\begin{align}
\label{intss}
   \langle \Phi,\, \Phi \rangle &= -i \int_{c}^1 \dd z \int_0^{2\pi} \dd \theta \sqrt{g_{\Sigma}(z)} \times n^{t} \hspace{1mm} \p{2i\omega} \hspace{1mm} \abs{R\p{z}}^2 \notag \\
    & \quad+i \int_c^1 \dd z \int_0^{2\pi} \dd \theta \sqrt{g_{\Sigma}(z)} \times  n^{\theta} \hspace{1mm} \p{2i \ell} \hspace{1mm} \abs{R(z)}^2
\end{align}
where
\begin{align}
   R(z)= \, _2F_1\left[-k,k+2 \alpha +\Delta ;2 \alpha +1;z\right]=\frac{\Gamma (2 \alpha +1) \Gamma (k+1)}{\Gamma (k+2 \alpha +1)}J_k^{(2 \alpha ,\Delta -1)}\left(1-2 z\right)
\end{align}
We have expressed the hypergeometric function as Jacobi polynomials since the first entry is a negative integer and the hypergeometric sum terminates\footnote{I thank David Grabovsky for pointing this out to me, who in turn was informed by Adolfo Holguin.}. Transforming the coordinate to $x=1-2z$, the first integral becomes (up to the ratios of gamma functions in the above transformation):
\begin{align}
    2\pi \omega_{\ell k}\cdot2^{-2\alpha-\Delta}\int_{-1}^{1-2c} \dd x \frac{1-2c-x}{1-x}\p{1-x}^{2\alpha}\p{1+x}^{\Delta-1}\abs{J_{k}^{\p{2\alpha,\Delta-1}}\p{x}}^2 
\end{align}
The Jacobi Polynomials obey the following identity: 
\begin{align}
    (1+x)^{2\alpha}(1+x)^{\Delta-1}J_{k}^{\p{2\alpha,\Delta-1}}\p{x}=\frac{\p{-1}^{k}}{2^k k!}\partial^k_x\Big[ (1-x)^{2\alpha+k}(1+x)^{k+\Delta-1}\Big]
\end{align}
For convenience, we define:
\begin{align}
    f(x)=\frac{1-2c-x}{1-x}, \quad g(x)=(1-x)^{2\alpha+k}(1+x)^{k+\Delta-1}
\end{align}
the integral now becomes:
\begin{align}
      \mathcal{F}_{k\ell}  \int_{-1}^{1-2c} \dd x \hspace{1.5mm} \partial^k_x\p{g(x)}f(x) J_{k}^{\p{2\alpha,\Delta-1}}\p{x}, \quad \mathcal{F}_{k\ell}=2\pi \omega_{\ell k}\cdot 2^{-2\alpha-\Delta} \cdot\frac{\p{-1}^{k}}{2^k k!}
\end{align}
Integrating by parts $k$-times, we end up with (up to $\mathcal{F}_{k\ell}$): 
\begin{align}
\label{integralss}
   \left. \sum_{s=2}^k \p{-1}^{s-1}\p{\partial_x^{k-s}g(x)}\partial_x^{s-1}\p{f(x)J_{k}^{\p{2\alpha,\Delta-1}}\p{x}} \right|^{1-2c}_{-1} +(-1)^k \int_{-1}^{1-2c}\dd x g(x)\partial_x^k\p{f(x)J_{k}^{\p{2\alpha,\Delta-1}}\p{x}}
\end{align}
Here the boundary terms evaluated at $x=-1$ will vanish due to the terms involving powers of $(1+x)$ in $g(x)$. The second derivative can be written as: 
\begin{align}
    \partial_x^{s-1}\p{f(x)J_{k}^{\p{2\alpha,\Delta-1}}\p{x}}&=\sum_{i=0}^{s-1}\frac{\Gamma\p{s}}{\Gamma\p{i+1}\Gamma\p{s-i}}f(x)^{\p{s-1-i}}\partial_x^{i}J_{k}^{\p{2\alpha,\Delta-1}}\p{x} \notag \\
    &=\sum_{i=0}^{s-1}\frac{\Gamma\p{s}\Gamma(s-i)}{\Gamma\p{i+1}\Gamma\p{s-i}}\p{-\frac{1}{\p{1-x}^{s-1-i}}+\frac{1-2c-x}{(1-x)^{s-i}}} \times \notag \\
   &  \hspace{6mm}\times \frac{\Gamma\p{k+i+2\alpha+\Delta}}{2^{i}\Gamma\p{k+\alpha+\Delta}}J_{k-i}^{\p{2\alpha+i,\Delta-1+i}}\p{x}
\end{align}
and the terms evaluated at $x=1-2c$ will not vanish. Combining everything, we have for the boundary term: 
\begin{align}
     & \left. \sum_{s=2}^k \p{-1}^{s-1}\p{\partial_x^{k-s}g(x)}\partial_x^{s-1}\p{f(x)J_{k}^{\p{2\alpha,\Delta-1}}\p{x}} \right|_{1-2c}= \\
    =&\sum_{s=2}^k \sum_{i=0}^{s-1} \p{-1}^{s}\p{\partial_x^{k-s}g(x)}\frac{\Gamma\p{s}\Gamma\p{k+i+2\alpha+\Delta}}{2^{i}\Gamma\p{i+1}\Gamma\p{k+\alpha+\Delta}}\p{\frac{1}{\p{1-x}^{s-1-i}}}\left. J_{k-i}^{\p{2\alpha+i,\Delta-1+i}}\p{x} \right|_{x=1-2c} \notag
\end{align}
which produces the result in \S.\ref{section3}. The second integral in (\ref{intss}) can be similarly evaluated, where we instead have: 
\begin{align}
    f(x)=\frac{1+x}{1-x}
\end{align}
Following the same procedure as before, we have: 
\begin{align}
    & \left. \sum_{s=2}^k \p{-1}^{s-1}\p{\partial_x^{k-s}g(x)}\partial_x^{s-1}\p{f(x)J_{k}^{\p{2\alpha,\Delta-1}}\p{x}} \right|_{1-2c}= \\
    =&\sum_{s=2}^k \sum_{i=0}^{s-1} \p{-1}^{s-1}\p{\partial_x^{k-s}g(x)}\frac{\Gamma\p{s}\Gamma\p{k+i+2\alpha+\Delta}}{2^{i}\Gamma\p{i+1}\Gamma\p{k+\alpha+\Delta}}\p{\frac{1}{\p{1-x}^{s-1-i}}}\left. J_{k-i}^{\p{2\alpha+i,\Delta-1+i}}\p{x} \right|_{x=1-2c} \notag
\end{align}
The boundary terms evaluated at $x=1-2c$ do not vanish as before. Note that the boundary terms for $I_1,I_2$ only differ by a sign, so one might expect for them to cancel. However, it is prevented by the prefactors in front of each integral.
\subsection{$-\beta_-^2\leq r^2 < \infty$}
In this region, $0\leq z <\infty$. With coordinate transformation $x=1-2z$, the first integral becomes (up to the ratios of gamma functions in the above transformation):
\begin{align}
    2\pi \omega_{\ell k}\cdot2^{-2\alpha-\Delta}\int_{-1}^{1}\frac{1-2c-x}{1-x}\p{1-x}^{2\alpha}\p{1+x}^{\Delta-1}\abs{J_{k}^{\p{2\alpha,\Delta-1}}\p{x}}^2
\end{align}
We can carry out the same analysis as above, and in this case, the boundary term no longer contributes. Since powers of $(1+x)$ and $(1-x)$ in $g(x)$ will make them vanish. The integral now follows from the orthogonality condition of the Jacobi polynomials: 
\begin{equation}
\int_{-1}^{1} (1 - x)^\alpha (1 + x)^\beta P^{(\alpha,\beta)}_m(x) P^{(\alpha,\beta)}_n(x) \, dx = \frac{2^{\alpha+\beta+1} \Gamma(n + \alpha + 1) \Gamma(n + \beta + 1)}{(2n + \alpha + \beta + 1)\Gamma(n + \alpha + \beta + 1) n!} \delta_{nm},
\end{equation}
where the first integral in (\ref{norm2}) can be evaluated as (including all the prefactors): 
\begin{align}
   I_1= \frac{\pi  \beta _+^2 \omega_{\ell k}  \Gamma (2 \alpha +1)^2 \Gamma (k+1) \Gamma (k+\Delta ) (2 \alpha -c (2 \alpha +\Delta +2 k))}{\alpha  \left(\beta _+^2-\beta _-^2\right) (2 \alpha +\Delta +2 k) \Gamma (k+2
   \alpha +1) \Gamma (k+2 \alpha +\Delta )}
\end{align}
Similarly, the second integral can be evaluated as: 
\begin{align}
   I_2= -\frac{\beta _- \beta _+ \ell \Gamma (2 \alpha +1)^2 (\Delta +2 k) \Gamma (k+1) \Gamma (k+\Delta )}{2 \alpha  \left(\beta _+^2-\beta _-^2\right) (2 \alpha +\Delta +2 k) \Gamma (k+2 \alpha +1) \Gamma (k+2
   \alpha +\Delta )}
\end{align}
Thus, combining the results, we have: 
\begin{align}
    \langle \Phi,\, \Phi \rangle &= \frac{\Gamma (2 \alpha +1)^2 \Gamma (k+1) \Gamma (k+\Delta )}{\Gamma (k+2 \alpha +1) \Gamma (k+2 \alpha +\Delta )}\mathcal{D}_{\ell k}, \notag \\
    \mathcal{D}_{\ell k}&=\frac{\beta _+ \bigg(\beta _- \ell (\Delta +2 k)-\beta _+ \omega_{k\ell}  \big(c (\Delta +2 (\alpha +k))-2 \alpha \big)\bigg)}{2 \alpha  \left(\beta _+^2-\beta _-^2\right) (\Delta +2\alpha +2k)}
\end{align}
which matches the results in \S.\ref{section3}.
\section{Geodesic length}
\label{Appen.B}
The geodesic length of the rotating particle solution is easily computed in the covering $\tilde{\text{AdS}}$, and then transformed to the quotient space \cite{Shenker:2013pqa,Maxfield:2014kra}. As discussed above, the AdS$_3$ spacetime can be constructed from flat $\mathbb{R}^{2,2}$ space:
\begin{align}
    \dd s^2=-\dd T_1^2-\dd T_2^2+\dd X_1^2+\dd X_2^2
\end{align}
Restricting to the submanifold $X^2_1+X^2_2-T_1^2-T_2^2=-1$, we obtain the geometry of AdS$_3$. The isometry group is $SO(2,2)$, and it can also be represented by $SL(2,\mathbb{R})\times SL(2,\mathbb{R})/\mathbb{Z}_2 $, with group element:
\begin{align}
  \bold{X}=  \left(
\begin{array}{cc}
 T_1+X_1 & \quad T_2+X_2 \\
 -T_2+X_2 & \quad T_1-X_1 \\
\end{array}
\right), \quad \text{det}|\bold{X}|=1
\end{align}
The spinning particle solution is obtained by the following identification: 
\begin{align}
    X_1&=\sqrt{\alpha}\cos\p{\beta_+\phi-\beta_-t}, &X_2=\sqrt{\alpha}\sin\p{\beta_+\phi-\beta_-t} \\
    T_1&=\sqrt{\alpha+1}\cos\p{\beta_+t-\beta_- \phi}, &T_2=\sqrt{\alpha+1}\sin\p{\beta_+t-\beta_-\phi}
\end{align}
where 
\begin{align}
    \alpha(r)=\frac{r^2+\beta_-^2}{\beta_+^2-\beta_-^2}, \quad 0\leq r<\infty
\end{align}
It is well known that in hyperbolic space, the length of geodesic between two points $\bold{X},\bold{X}'\in SO(2,2)$ is given by: 
\begin{align}
   d=\cosh^{-1}\p{\frac{\bold{X}^{-1}\bold{X}'}{2}}
\end{align}
This gives: 
\begin{align}
    d=\cosh^{-1}\p{T_1T'_1+T_2T'_2-X_1X_1'-X_2X_2'}
\end{align}
Choosing $T_i'\p{t=0,\phi=0},X_i'(t=0,\phi=0)$, taking the limit $r\rightarrow\infty$, the normalized geodesic length can be easily obtained as: 
\begin{align}
    d=\ln\Big[\frac{2}{\beta_+^2-\beta_-^2}\Big(\cos\p{\beta_+t-\beta_-\phi}-\cos\p{\beta_-t-\beta_+\phi}\Big)\Big]
\end{align}
This precisely produces the bulk propagator (\ref{propagator1}). Such geodesic computation is not sensitive to the region containing CTCs since it does not depend on the behavior near $r=0$. More explicitly, we can compute its turning point. It is easiest if we restrict to the constant time-slice, where we have: 
\begin{equation}
    \dd s^2_{\Sigma}=\p{\frac{1}{n^2}+r^2+\frac{J^2}{4r^2}}^{-1}\dd r^2+r^2 \dd \phi^2
\end{equation}
In this case, we have one KVF: $\partial_{\phi}$, giving rise the conserved angular mometum $L=r^2\frac{d\phi}{ds}$. Thus, the geodesic equation can be obtained as: 
\begin{align}
    \Dot{r}^2=\p{\frac{1}{n^2}+r^2+\frac{J^2}{4r^2}}\p{1-\frac{L^2}{r^2}}
\end{align}
and the only acceptable turning point is at $r=L>0$. Thus, this spacelike geodesic does not probe into the region containing CTCs.

\section{The Mode Sum}
\label{Appen.C}
Here we show that the mode sum (\ref{finalpropagator}) is equivalent to the expression (\ref{moi}). Instead of directly evaluating the sum, we adopt the spectral decomposition approach discussed in \cite{Hijano:2019qmi,Grabovsky:2024jwf}, where by analytically continuing the trigonometry expression (\ref{moi}), we obtain the mode sum expression. Note that we can write (\ref{moi}) in terms of complex coordinates $z=e^{\tau+i \phi}$ and $\bar{z}=e^{\tau-i \phi}$, where $i\tau=t$. The propagator (\ref{moi}) is then:
\begin{align}
    G^{+}\p{z,\bar{z}} = \p{\alpha_H \bar{\alpha}_H}^{\Delta} \sum_{m=-\infty}^{+\infty} z^{\alpha_H \cdot\frac{\Delta}{2}} \bar{z}^{\bar{\alpha}_H \cdot \frac{\Delta}{2}}e^{2\pi i m \Delta \beta_- }\p{1-z^{\alpha_H} e^{2\pi i m \alpha_H}}^{-\Delta}\p{1-\bar{z}^{\bar{\alpha}_H} e^{-2\pi i m \bar{\alpha}_H}}^{-\Delta}
\end{align}
The binomial theorem tells us: 
\begin{align}
    \p{1-z^{\alpha_H} e^{2\pi i m \alpha_H}}^{-\Delta}&=\frac{1}{\Gamma\p{\Delta}}\sum_{p=0}^{\infty}\frac{\Gamma\p{\Delta+p}}{\Gamma\p{p+1}}\p{z^{\alpha_H} e^{2\pi i m \alpha_H}}^p \notag \\
    \p{1-\bar{z}^{\bar{\alpha}_H} e^{-2\pi i m \bar{\alpha}_H}}^{-\Delta}&=\frac{1}{\Gamma\p{\Delta}}\sum_{q=0}^{\infty}\frac{\Gamma\p{\Delta+q}}{\Gamma\p{q+1}}\p{\bar{z}^{\bar{\alpha}_H} e^{-2\pi i m \bar{\alpha}_H}}^q
\end{align}
Performing the sum over $m$ yields the Dirac comb: 
\begin{align}
    \sum_{m=-\infty}^{\infty} \exp{2\pi i m \p{p \cdot \alpha_H -q\cdot \bar{\alpha}_H+\Delta \beta_-}}=\sum_{\ell=-\infty}^{\infty}\delta\big(p \cdot \alpha_H -q\cdot \bar{\alpha}_H+\Delta \beta_- -\ell \big)
\end{align}
This effectively imposes the condition: 
\begin{align}
    p \cdot \alpha_H -q\cdot \bar{\alpha}_H+\Delta \beta_- -\ell =0
\end{align}
The delta function is infinite when the above condition is satisfied for each integer value of $p$. Schematically, we have:
\begin{align}
    \sum_{\ell,q}\sum_{p}\delta\p{0}_p \times \p{\frac{1}{\G \p{\Delta}^2}\frac{\G\p{\Delta+q}}{\G\p{q+1}}\cdot \frac{\G\p{\Delta+p}}{\G\p{p+1}}}
\end{align}
In order to extract the finite piece,  we should normalize it term by term by dividing out the infinities from the delta function. We shall only keep the finite piece and discard the sum over $p$ from now on.

Setting $q=k$\footnote{setting $p=k$ would correspond to the other quantization condition differing by a sign.}, and remembering that $\alpha_H=\beta_+ +\beta_-,\bar{\alpha}_H=\beta_+ -\beta_-$, we can then solve for $p$:
\begin{align}
    p=\frac{\ell-(2k+\Delta)\beta_-}{\beta_+ +\beta_-}+k
\end{align}
This is nothing but $p=2\alpha+k$ if one chooses the quantization condition (\ref{choosestuff}). With this quantization condition at hand, as well as $p=2\alpha+k, \hspace{1mm}q=k$, we have:
\begin{align}
   z^{\alpha_H \cdot\frac{\Delta}{2}} \bar{z}^{\bar{\alpha}_H \cdot \frac{\Delta}{2}} z^{p\cdot\alpha_H} \bar{z}^{q\cdot\bar{\alpha}_H}=e^{-i\omega_{k\ell}t}e^{i\ell \phi}
\end{align}
where we have transformed back to the $(t,\phi)$ coordinates, and $\omega_{k \ell }$ is given by (\ref{choosestuff}). Now, combining everything, the finite piece of the propagator can be expressed as: 
\begin{align}
     G^{+}\p{z,\bar{z}} = \p{\beta_+^2-\beta_-^2}^{\Delta} \sum_{k=0}^{\infty}\sum_{\ell=-\infty}^{\infty}\frac{\Gamma\p{\Delta+k+2\alpha}\Gamma\p{\Delta+k}}{\Gamma\p{\Delta}^2\Gamma\p{1+k+2\alpha}\Gamma\p{1+k}} e^{-i\omega_{k\ell}t}e^{i\ell \phi}
\end{align}
Thus, the mode sum (\ref{finalpropagator}) is equivalent to the expression (\ref{moi}) up to appropriate normalization and an irrelevant factor $\p{\beta_+^2-\beta_-^2}^\Delta\p{\beta_+ +\beta_-}$.

\subsection{The Finite Sum}
The mode sum can be expressed as a finite sum over image geodesics if the following conditions are satisfied:
\begin{align}
\label{conditionss}
    \alpha_H^{-1}&=\kappa \cdot n \in \mathbb{Z}_+, \quad n \in \mathbb{Z}_+ , \kappa\in \mathbb{Q} \notag \\
    \bar{\alpha}_H^{-1}&=\frac{\kappa}{s}\cdot n, \hspace{10mm}\quad s\in \mathbb{Z}_{+\p{\text{odd}}} \notag \\
    1\leq \hspace{1mm}&s< \sqrt{3}\kappa
\end{align}
where we will derive them in the next subsection. One can then show that the mode sum (\ref{finalpropagator}) is equivalent to: 
\begin{align}
\label{moi2}
    G^{+}\p{t,\phi}=\sum_{m=0}^{\alpha_H^{-1}-1} \bigg[ \frac{4}{\beta_+^2-\beta_-^2}\sin\p{\frac{\beta_++\beta_-}{2}\p{\phi+2\pi m-t}}\sin\p{\frac{\beta_+-\beta_-}{2}\p{\phi+2\pi m+t}} \bigg]^{-\Delta}
\end{align}
We can similarly apply the spectral decomposition method to the above expression. Everything is unchanged from before until we reach the step to perform the sum over $m$, where we now have: 
\begin{align}
    \sum_{m=0}^{\alpha_H^{-1}-1} \exp{2\pi i m \frac{\p{p -q\cdot \frac{\bar{\alpha}_H}{\alpha_H}+\frac{\Delta \beta_-}{\alpha_H}}}{\alpha_H^{-1}}}=\alpha_H^{-1}\sum_{\ell=-\infty}^{\infty}\delta\bigg[\p{p -q\cdot \frac{\bar{\alpha}_H}{\alpha_H}+\frac{\Delta \beta_-}{\alpha_H}}- \alpha_H^{-1} \ell \bigg]
\end{align}
where $\sum_k\delta \left[x-k n\right]$ is the Kronecker comb. Summing over $p$ imposes the condition: 
\begin{align}
    p \cdot \alpha_H -q\cdot \bar{\alpha}_H+\Delta \beta_- -\ell =0
\end{align}
and the rest follows as before. Note that in this case we no longer have infinite contributions due to the discrete nature of the Kronecker comb. Setting $J=0$, $\alpha_H=\bar{\alpha}_H=1/n$, the above sum precisely reduces to the form of a non-spinning defect.

\subsection{Derivation of \ref{conditionss}}
As discussed before, in case of vanishing angular momentum, the mode sum can only be written as a finite sum over image geodesics if the defect mass $1/n^2$ satisfies the condition $n \in \mathbb{Z}_+$. Here we show that certain conditions have to be satisfied for both the mass and angular momentum of the defect in order to write the propagator as a finite sum over images. It is clear that the mode sum obeys the periodicity of $\phi \sim \phi+2\pi$, and the point of summing over image geodesics is to restore this periodicity in the propagator. Thus, we must have: 
\begin{align}
    G^+(\phi)=G^+(\phi+2\pi)
\end{align}
In terms of the method of images, this effectively imposes the condition that the last image geodesic should become the minimal one under $\phi \sim \phi+2\pi$: 
\begin{align}
    \sin\p{\frac{\alpha_H}{2}\p{\phi+2\pi m}}\sin\p{\frac{\bar{\alpha}_H}{2}\p{\phi+2\pi m}}\xrightarrow{\phi \sim \phi+2\pi}\sin\p{\frac{\alpha_H}{2}\phi}\sin\p{\frac{\bar{\alpha}_H}{2}\phi}
\end{align}
where $m=\alpha_H^{-1}-1$ represents the last image geodesic. This condition can be satisfied if we choose (\ref{conditionss}), where the left-hand side becomes: 
\begin{align}
    \sin\p{\frac{\phi+2\pi \p{\kappa n-1}}{2\kappa n}}\sin\p{\frac{\phi+2\pi \p{\kappa n-1}}{2\kappa n}\cdot s}\xrightarrow{\phi \sim \phi+2\pi} (-1)^{s+1}\sin\p{\frac{\phi}{2\kappa n}}\sin\p{\frac{\phi \cdot s}{2\kappa n}}
\end{align}
Indeed the right-hand side expression is that of the minimal geodesics if $s$ is a positive odd integer. The inequality satisfied by $s$ can be derived from the non-extremal nature of the defect. Using (\ref{conditionss}), the angular momentum can be solved as: 
\begin{align}
    J=\frac{1}{2}\p{\frac{s^2}{\kappa^2}-1}\frac{1}{n^2}
\end{align}
Since we want $\abs{J}<1/n^2$, this yields $s<\sqrt{3}\kappa$.

\bibliographystyle{jhep}
\bibliography{refs}

\end{document}